\documentclass[manuscript]{aastex}

\shorttitle{YSO inner disk structures}
\shortauthors{S. Ragland et al.}
\usepackage{graphicx}
\usepackage[mathscr]{eucal}
\begin{document}

\title{First L-band Interferometric Observations of a Young Stellar Object: Probing the Circumstellar Environment of MWC~419}
\author{S. Ragland\altaffilmark{1},
R. L. Akeson\altaffilmark{2},
T. Armandroff\altaffilmark{1},
M. M. Colavita\altaffilmark{3},
W. C. Danchi\altaffilmark{4},
L. A. Hillenbrand\altaffilmark{5},
R. Millan-Gabet\altaffilmark{2},
S. T. Ridgway\altaffilmark{6},
W. A. Traub\altaffilmark{3},
P. L. Wizinowich\altaffilmark{1}
}

\altaffiltext{1}{W. M. Keck Observatory, 65-1120 Mamalahoa Hwy, Kamuela, HI 96743; sragland@keck.hawaii.edu}
\altaffiltext{2}{NExScI, California Institute of Technology, 770 South Wilson Avenue, Pasadena, CA 91125}
\altaffiltext{3}{Jet Propulsion Laboratory,  California Institute of Technology, M/S 301-451, 4800 Oak Grove Dr., Pasadena CA, 91109}
\altaffiltext{4}{NASA Goddard Space Flight Center, Exoplanets \& Stellar Astrophysics, Code 667, Greenbelt, MD 20771}
\altaffiltext{5}{California Institute of Technology, Pasadena, CA 91125}
\altaffiltext{6}{National Optical Astronomy Observatories, P.O. Box 26732, Tucson, AZ 85726-6732} 

\begin{abstract}
We present spatially-resolved K- and L-band spectra 
(at spectral resolution R = 230 and R = 60, respectively) 
of MWC 419, a Herbig Ae/Be star.  The data were obtained simultaneously with a new configuration of the 85-m baseline Keck Interferometer.  Our observations are sensitive to the radial distribution of temperature in the inner region of the disk of MWC 419. We fit the visibility data with both simple geometric and more physical disk models.  The geometric models (uniform disk and Gaussian) show that the apparent size increases linearly with wavelength in the 2-4 $\mu$m wavelength region, suggesting that the disk is extended with a temperature gradient. A model having a power-law temperature gradient with radius simultaneously fits our interferometric measurements and the spectral energy distribution data from the literature. The slope of the power-law is close to that expected from an optically thick disk. 
Our spectrally dispersed
interferometric measurements include the Br $\gamma$ emission line. 
The measured disk size at and around Br $\gamma$ suggests that emitting 
hydrogen gas is located inside (or within the inner regions) of the dust disk. 

\end{abstract}

\maketitle

\section{Introduction}

Pre-main sequence (PMS) stars fall to the upper right of the main sequence on the Hertzsprung-Russell (HR) diagram, 
journeying towards the main sequence via radial contraction.
For the first several million years, a PMS star is surrounded by a disk of gas and dust, 
left over from the 
early stage of star formation. 

The evolution of a disk is not well understood. 
Circumstellar disks provide the raw material for planet formation. 
Thus, understanding the evolution of a disk helps us understand planet formation.
Clues to the physical conditions of planet formation and hence future suitability for life on planets other than Earth potentially can be inferred 
from a detailed characterization of inner young stellar object (YSO) disks. 

Herbig Ae/Be (HAeBe) stars are intermediate-mass PMS stars with spectral types earlier than G0, located close to the zero-age main-sequence. 
Early spectral energy distribution (SED) modeling efforts \citep{1992ApJ...397..613H} to explain the infrared excess of HAeBe stars assumed (a) optically thick but geometrically thin circumstellar disks with inner optically thin holes, up to several stellar radii, in order to account for the observed inflections in the 1-5$\mu$m region, and (b) relatively large accretion rates, on the order of 10$^{-6}$ M$_\odot$ yr$^{-1}$, in order to account for the strength of the 3.8$\mu$m (L-band) excess.  It was quickly pointed out \citep{1993ApJ...407..219H} that at such a large mass accretion rate, the gas in the inner dust-free hole is expected to be optically-thick and to produce excess near-infrared emission, which is inconsistent with the dip seen in the near-infrared SED. 
For the relatively low accretion rates later inferred for HAeBe objects, disk models could not explain the observed SEDs in the near- infrared region. Puffed-up inner dust-rim models were introduced \citep{Natta01} and later refined \citep{Dullemond01,2005A&A...438..899I} to ameliorate the shortcomings of the classical disk models. These later models attribute the near-infrared excess to stellar radiation shining directly on the inner dust edge of the dust disk. Tuthill et al. (2001) directly imaged the high luminosity YSO LkH$\alpha$~101 and independently proposed that the bulk of the near-infrared emission arises in a hot ring located at the dust sublimation radius. 

Long-baseline interferometric observations provide the milliarcsecond angular resolution required 
to resolve the planet-forming structures immediately surrounding PMS stars.  Such
high-angular-resolution infrared observations are well suited for probing near-circumstellar environments since the 
inner zones of the circumstellar disks and envelopes emit primarily at near-infrared wavelengths. A large number of YSOs across the luminosity range have now been spatially resolved at near-infrared wavelengths \citep{Millan-Gabet01, Colavita03, 2003ApJ...588..360E, 2004ApJ...613.1049E, 2007ApJ...657..347E, 2005ApJ...624..832M, Akeson05, 2007prpl.conf..539M, 2008A&A...483L..13I, 2009ApJ...692..309E, Tannirkulam08}, showing that indeed the characteristic sizes correlate strongly with central luminosity, lending support to the ``puffed-up'' inner dust rim paradigm, especially for Herbig Ae objects. However, it was also pointed out \citep{2002ApJ...579..694M, 2004ApJ...613.1049E, 2005ApJ...624..832M} that most higher luminosity Herbig Be objects are considerably undersized compared to the predictions of this model, and in better agreement with the ``classical'' models featuring optically thick emission that extends inward very close to the central star.

MWC~419 is a photometrically variable emission line star that is in many ways
typical of the HAeBe class, including illumination of a reflection nebula \citep{1960ApJS....4..337H}.  
MWC~419 has spectral type B8 \citep{1960ApJS....4..337H}, luminosity 330 L$_{\odot}$ \citep{1992ApJ...397..613H}, and distance 650 pc \citep{1992ApJ...397..613H}; 
it is also known as V594 Cas or BD+61 154. 
Narrow band 
H$\alpha$ imaging studies show unipolar large scale structures around MWC~419 \citep{2008A&A...477..193M}. Interestingly, these 
images show two lobes on the south-east side of MWC~419 - the outer lobe extending to $\sim$ 4 arcminutes - suggesting a history 
of episodic mass-loss in this PMS star's evolution. These authors attribute the non-detection of north-west lobes to obscuration. These large scale structures have no influence on our observations of the central region.
The measured intrinsic polarization of MWC~419 in the V-band is 0.53\% \citep{1992ApJ...397..613H} 
suggesting that the contributions from reflected optical light to the total emission is insignificant.
No close companions to MWC~419 are known \citep{1997ApJ...481..392P, 1993A&A...278L..47B}.
The star has P-Cygni characteristics in the lower Balmer lines and the implied wind has been modeled in detail by
Bouret \& Catala (1998). 
However, it has also been considered as a B[e] star and interpreted as undergoing post-main sequence evolution 
rather than pre-main sequence evolution.  The star is projected on the outer 
regions of the young association NGC 225 \citep{1960ApJS....4..337H, 2006BASI...34..315S}, within a small area 
of enhanced extinction and CO emission \citep{1995PhDT.........1H, 2002A&A...387..977F}
that is associated with the dark cloud L~1302 \citep{1998A&A...330..145V}. 
However, the membership of the star in the cluster is debated (Lattanzi et al. 1991).
The age of the cluster is often quoted as 120 Myr \citep{1991AJ....102..177L} but Subramaniam et al. (2006)
have argued based on the
presence of a number of young stars that it is really a $<$10 Myr pre-main sequence population.

As an IRAS and ISO source, the infrared dust spectrum and spectral energy distribution of MWC~419 
have been discussed by several authors \citep{1993AJ....106..656B, 2000Ap&SS.271..259C, 1992ApJ...398..254B, 1992ApJ...397..613H, 1997ApJ...475L..41M, 1997ApJ...485..290P}. No Spitzer data were obtained for this source but  
it has a weak silicate emission at 10 $\mu$m in ground-based data \citep{2000Ap&SS.271..259C}.

MWC~419 was previously observed interferometrically using the Palomar Testbed Interferometer (PTI), and was spatially resolved in the K-band with a reported uniform-disk angular diameter of 3.34 $\pm$0.16 milliarcsec (mas), or a Gaussian distribution FWHM of 2.07$\pm$0.11 mas \citep{2003Ap&SS.286..145W}. MWC~419 was not resolved from the Infrared Optical Telescope Array (IOTA) interferometer observations \citep{Millan-Gabet01}, where the angular resolution was $\sim$ 6 mas at 2.2$\mu$m.
  
The Keck Interferometer (KI), with $\sim$3 mas angular resolution at 2.2$\mu$m, can resolve the inner disks ($\sim$ 1 AU) 
of nearby PMS stars.  In this paper we report observations of MWC~419 using multi-color interferometry at well-separated wavelengths, here K-band (2.0-2.4 $\mu$m) and L-band (3.5-4.1 $\mu$m). The simultaneous K- and L-band interferometric measurements enable us to probe different regions 
($\sim$ 1300K and 800K respectively) of the inner circumstellar disks of PMS stars. 
Discrete spatial distributions, such as 
dust-rims, and relatively-smooth spatial distributions, such as classical accretion disks, are expected to have
different size vs. wavelength behaviors and can be distinguished in such multi-color observations at well separated wavelengths. In addition, 
interferometric measurements in the relatively unexplored L-band provide further constraints to the 
disk/envelope geometry via temperature- and density-sensitivity of different models. Multicolor interferometry at well separated wavelengths (H, K and N-bands) was recently reported for two other YSO disks, namely, MWC~297 \citep{2008A&A...485..209A} and MWC~147 \citep{2008ApJ...676..490K} from VLTI observations.

In Section 2, below, we present our observations and data reduction. In Section 3 we describe the data analysis where we fit the visibility-squared ($V^2$) data with various YSO disk models. In Section 4 
we discuss our results and in Section 5 we provide a brief summary.

\section{Observations and Data Reduction}
KI is a near- and mid-infrared long-baseline interferometer consisting of two 10-m diameter apertures 
separated by a $B$ = 85-m baseline at a position angle of $\sim$ 38$^o$ east of north. Both Keck telescopes are equipped with 
adaptive optics.
The maximum resolution of our KI observations is $\lambda$/2$B$ $\sim$2.7 mas and $\sim$4.5 mas 
in the K-band and L-band respectively. 

The L-band science instrument, consisting of an infrared camera and beam-combiner optics, is similar to the existing H- and K-band instrument (FATCAT) routinely used for $V^2$ measurements since 2002 \citep{Colavita03}. We obtained first light with this new instrument in April 2008. The L-band limiting magnitude of 6 is set by the requirement for a broadband phase signal-to-noise ratio (SNR) of 10 in the 10 ms fringe-tracker frame.  The total integration time on source is typically 200 sec, which provides good SNR for $V^2$ measurements in the 10 spectral channels. 

The L-band instrument uses a PICNIC focal plane array detector with a 5 $\mu$m cutoff wavelength, while FATCAT uses a HAWAII array. The detector electronics are similar for both instruments. 
Details of this L-band instrument are given in Ragland et al. (2008).



For the current work a prototype multi-wavelength observing capability
was used, enabling simultaneous K- and L-band observations of our science target.
In this configuration, the telescope pupil is split into left and right-halves at the dual-star modules of both telescopes and routed through separate coude paths utilizing the existing beam-train infrastructure for the nulling mode. The left-half pupils of both telescopes are combined by the K-band instrument and the right-half pupils are combined by L-band instrument.

The K- and L-band science instruments each have 2 complementary interferometric outputs with different spectrometers on each.  For L-band, these are a (pseudo) broad-band, and a low dispersed mode (10 channels across the L-band; R = 60). For K band, these are a broad-band, and a medium dispersion mode (42 channels across the K-band; R = 230).

The field-of-view of the instrument, defined by the single mode fibers that couple that light to the detector array, is $\sim$ 55 mas for the K-band and $\sim$ 93 mas for the L-band observations. These field restrictions were enforced in our modeling work. 

The ZABCD algorithm \citep{Colavita03} is used for fringe tracking and science measurements. In this procedure five reads are made while the fast delay line scans over one wavelength (2.2 $\mu$m for the K-band and 3.7 $\mu$m for the L-band); explicitly, the reads are the reset pedestal (z), followed by four non-destructive reads (a, b, c, d) spaced at $\lambda$/4 intervals as the detector integrates up. For each quarter-wave bin we then calculate $A = a - z$, $B = b - a$, etc. 
These values are used to estimate the square of the fringe visibility ($V^2$) for science, and the fringe phase and SNR for fringe tracking.  We apply corrections to the bin data to account for detector bias and for differences between the length of the delay scan and the actual wavelength. Residual instrument bias and atmospheric seeing effects are corrected by observing a calibrator of known $V^2$ under similar observing conditions. Essentially, we estimate instrument transfer function using the bracketing observations of calibrator stars with known angular diameters using a weighted averaging scheme that considers the time and sky proximity of calibrators relative to the target\tablenotemark{1}. \tablenotetext{1}{http://nexsci.caltech.edu/software/V2calib/}

The observations reported here were taken on the night of UT 19 August 2008. We observed two calibrators - HD~1843 and HD~6210 - under similar observing conditions as the science target to calibrate the 
science data. The adopted angular sizes of the calibrators are 1.0$\pm$0.02 and 0.5$\pm$0.01 mas respectively \citep{1999PASP..111.1515V}. We performed bracketed calibration meaning that our observing sequence consisted of calibrator-target-calibrator measurements. 

The measurements presented here were taken over a narrow range of position angles (25$^o$ to 33$^o$) and projected baselines (76.1m to 77.8m), and hence no attempt was made to derive an inclination angle and position angle of the disk. For the purpose of this article, we assume that the disk is face-on. 
Thus, the sizes reported here are the values along the position angle of about 29$^o$, and the actual size of the disk could be larger depending on the inclination angle and position angle. 
Our K-band measurements, taken at the position angle of about 29$^o$, gives similar disk size as that of previous PTI observations, taken at the position angle of about 83$^o$. Hence, a face-on disk for MWC~419 is a resonable assumption in the absence of necessary high spatial resolution observations. 

\section{Analysis}
In this section, we fit our measurements with different models of increasing complexity and increasing physical realism. 

The measured $V^2$ includes contributions from the central star  (V$^2_{*}$) and the bright circumstellar disk. The visibility-square (V$^2_{disk}$) of the circumstellar disk of MWC~419 is obtained from the measured data by removing the contributions from the 
central star \citep{Millan-Gabet01} using

\begin{equation}
V^2_{measured} = \left(\frac{F_*V_*+F_{disk} V_{disk}}{F_*+F_{disk}}\right)^2
\label{star_disk}
\end{equation}

The adopted disk-to-star flux ratio (F$_{disk}$/F$_*$) is 12 and 40 in K and L based on our SED analysis discussed 
in Section 3.3. The central star is assumed to be unresolved for our observations (i.e., V$_*$ = 1.0), which is a reasonable 
assumption for a B8 star at a distance of 650pc. These corrections yield values of $V_{disk}^2$ that are smaller than the total $V^2$ by 0.04 ($\sim$ 9\%) and 0.01 ($\sim$ 2\%) in K and L respectively. The K-band $V^2$ measurements are consistent with earlier K broadband measurements and provide additional spectrally resolved information within the K-band spectral region. 

The visibilities of the disk models (Section 3.2 and 3.3) are computed by numerically summing the contributions from annular rings of infinitesimally small widths and weighting them by their respective flux contributions; for this purpose, we divided the disk radially into 5000 annular rings, in logarithmic scale. The normalized visibility for a uniformly bright annular ring with an inner diameter of $\theta_{in}$, an outer diameter of $\theta_{out}$ and $\theta_{in} \lesssim \theta \lesssim \theta_{out}$, can be written as
  
\begin{equation}
V_{ann}(B, \lambda, \theta) = \left(\frac{2\lambda}{\pi B (\theta_{out}^2-\theta_{in}^2)}\right) \left(\theta_{out} J_1(\pi B \theta_{out}/\lambda) - \theta_{in} J_1(\pi B \theta_{in}/\lambda)\right)
\label{vdisk}
\end{equation}
where $J_1$ is the first-order Bessel function, $B$ is the projected baseline and $\lambda$ is the wavelength of observation. 

The flux from such an annular uniform ring can be written as 

\begin{equation}
F_{ann}(\lambda, \theta) = \frac{\pi}{4} ~ P(T,\lambda) ~ (\theta_{out}^2 - \theta_{in}^2)
\label{fdisk}
\end{equation}
where $P(T,\lambda)$ is the Planck blackbody function and $T$ is the mean temperature of the annular ring. 

Now, the visibility of the total disk can be written as

\begin{equation}
V_{disk}(B, \lambda) = \sum{\left(V_{ann}(B, \lambda, \theta) \times \frac{F_{ann}(\lambda, \theta)}{\sum F_{ann}(\lambda, \theta)}\right)}
\label{vdisk2}
\end{equation}
where $\sum F_{ann}(\lambda, \theta)$ is the total disk flux derived by summing all the annular rings of the disk in order to normalize the flux. We adopted a nonlinear least-squares fitting method to minimize the chi-square for all our modeling work. 

For our SED analysis, photometric measurements from the literature are used, after interstellar extinction corrections using the extinction law of Cardelli et al. (1989).
The extinction for MWC~419 in the V-band is assumed to be $A_v$ = 2.1 \citep{1992ApJ...397..613H}. 

The models presented in this article assume face-on geometry for the
disk. A disk with an inclination angle $\phi$ could predict a larger
inner disk radius depending on the inclination angle and the position angle of the projected
interferometer baseline on the sky. In this case, SED models would
underestimate the flux at all wavelengths by a factor of
$cos(\phi)$. In other words, classical accretion disk models would
have a factor of $cos(\phi)$ larger accretion rate to explain the SED
data and power-law models would have a relatively larger disk
temperature at all radial distances. 

\subsection{Simple wavelength-dependent geometrical models}
The measurements presented in this article consist of spectrally dispersed data within the K- and L-bands. 
As stated above, the K and L visibilities corrected for the stellar contribution are different, suggesting a wavelength dependence 
to the size of the spatially resolved emission.   We can use the spectrally
dispersed data to investigate in more detail this wavelength dependency.
We choose three geometrical models, namely, uniform-disk, Gaussian distribution and ring models, to fit to our measurements.

The complex visibilities for a normalized pole-on uniform-disk and a Gaussian distribution
can be written as follows: 

\begin{equation}
V_{UD}(B, \lambda) = \frac{2 J_1[\pi B \theta_{UD}/\lambda]}{\pi B \theta_{UD}/\lambda}
\label{ud_eqn}
\end{equation}

\begin{equation}
V_{Gaussian}(B, \lambda) = exp\left(-\frac{(\pi B \theta_{FWHM}/\lambda)^2}{4~ln~2}\right)
\label{gaussian_eqn}
\end{equation}
Here $\theta_{UD}$ is the uniform-disk angular diameter of the circumstellar disk, $\theta_{FWHM}$ is the full-width at half maximum (FWHM) of the Gaussian distribution, and other terms as above. 
See Eqn.~\ref{vdisk} for the ring model.

\begin{figure}[bthp]
\centering
\includegraphics[width=0.475\hsize]{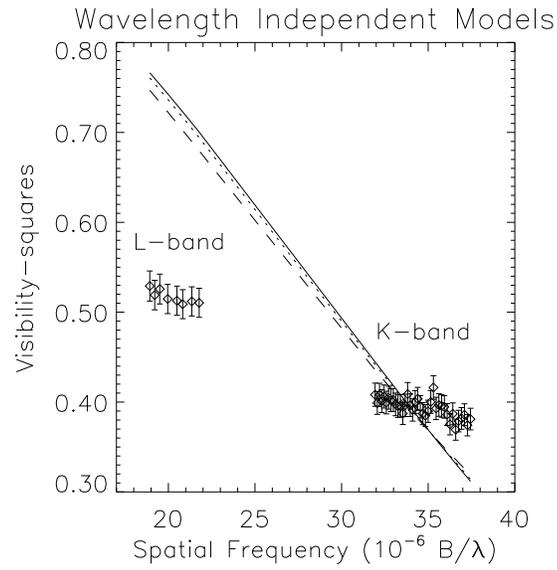}
\caption{Wavelength-independent model fits. The multi-wavelength measurements (diamond symbols) are plotted against spatial frequency. Also shown are geometrical model fits to the data; the dotted-line, dashed-line and continuous line refer to uniform-disk, Gaussian and ring models respectively. The poor model fits indicate that these simple models are an inadequate description of the data.
}
\label{basic}
\end{figure}

To start, we fit the data with a wavelength-independent geometrical model (Figure~\ref{basic}). The derived uniform-disk, Gaussian and ring angular sizes, from simultaneous fits to K and L band data, are 3.59 $\pm$ 0.01 mas, 2.21 $\pm$ 0.01 mas (FWHM) and 3.91 $\pm$ 0.01 mas respectively. The corresponding reduced-chi-square ($\chi^2_R$) values are 36.1, 35.0 and 38.3 respectively. As indicated by the very poor $\chi^2_R$ values, the uniform disk, Gaussian and ring models fit neither the K and L data in the mean, nor the slope of the visibilities throughout either the K- or L-bands. 

\begin{figure}[bthp]
\centering
\includegraphics[width=0.32\hsize]{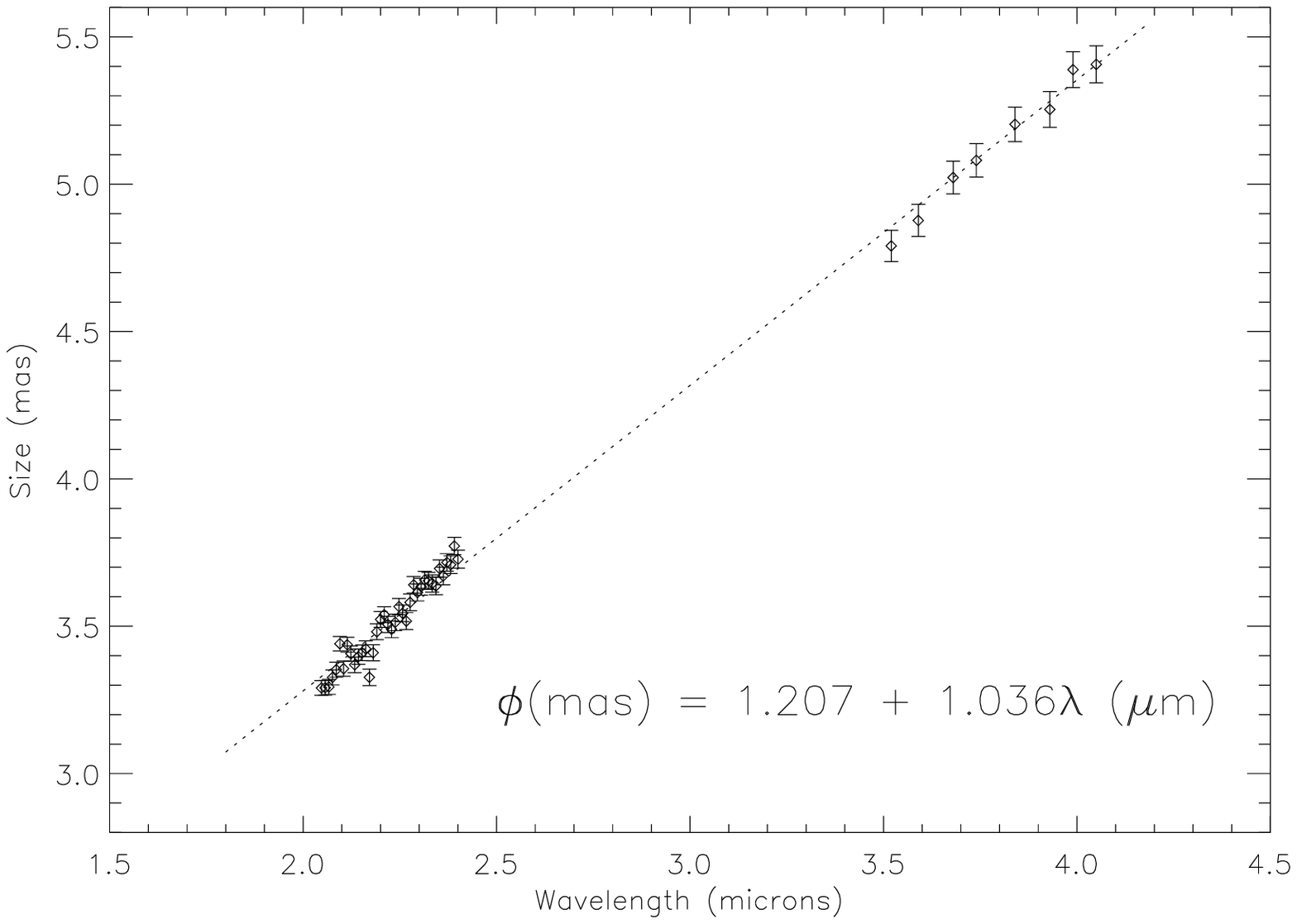}
\includegraphics[width=0.32\hsize]{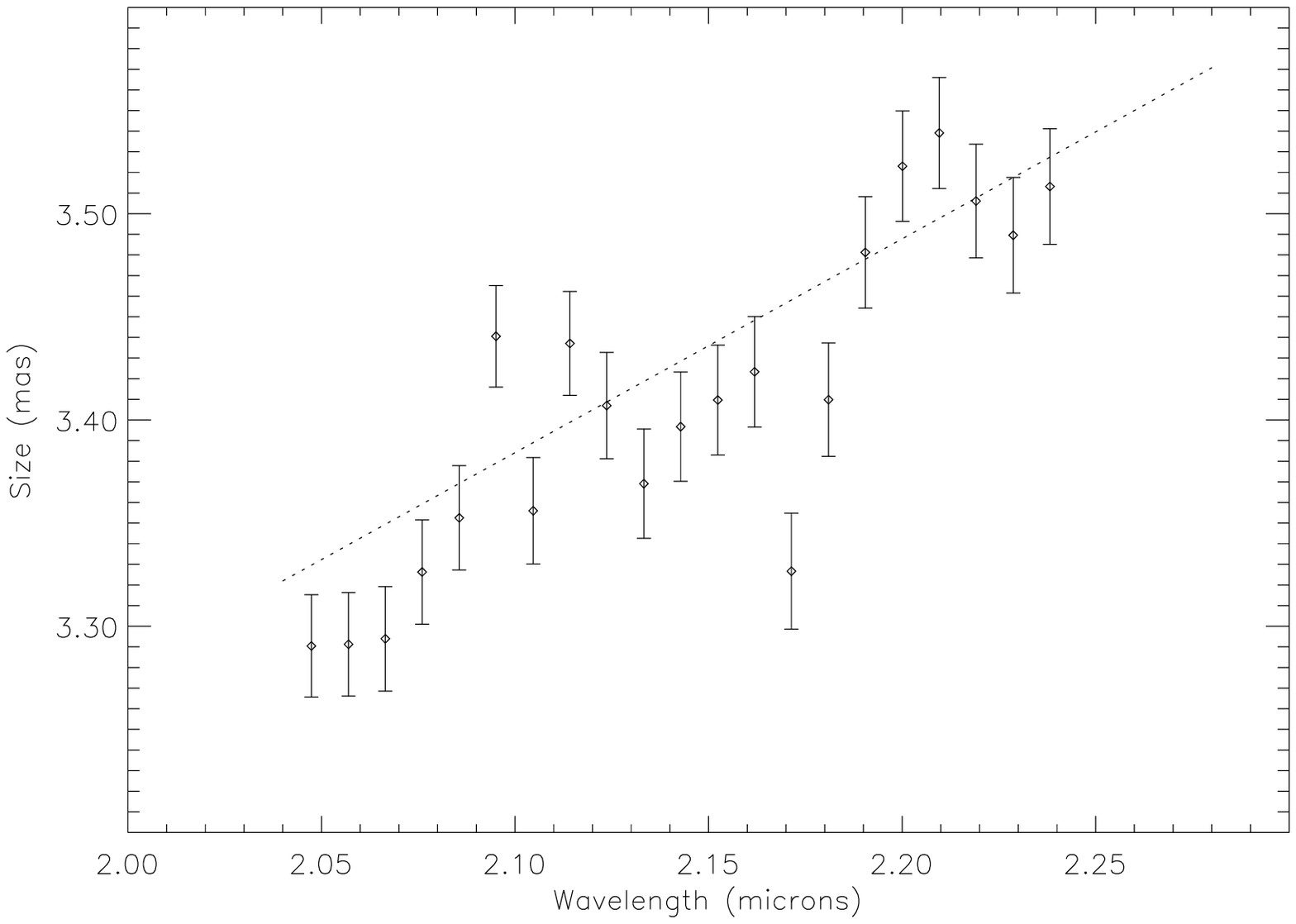}
\includegraphics[width=0.32\hsize]{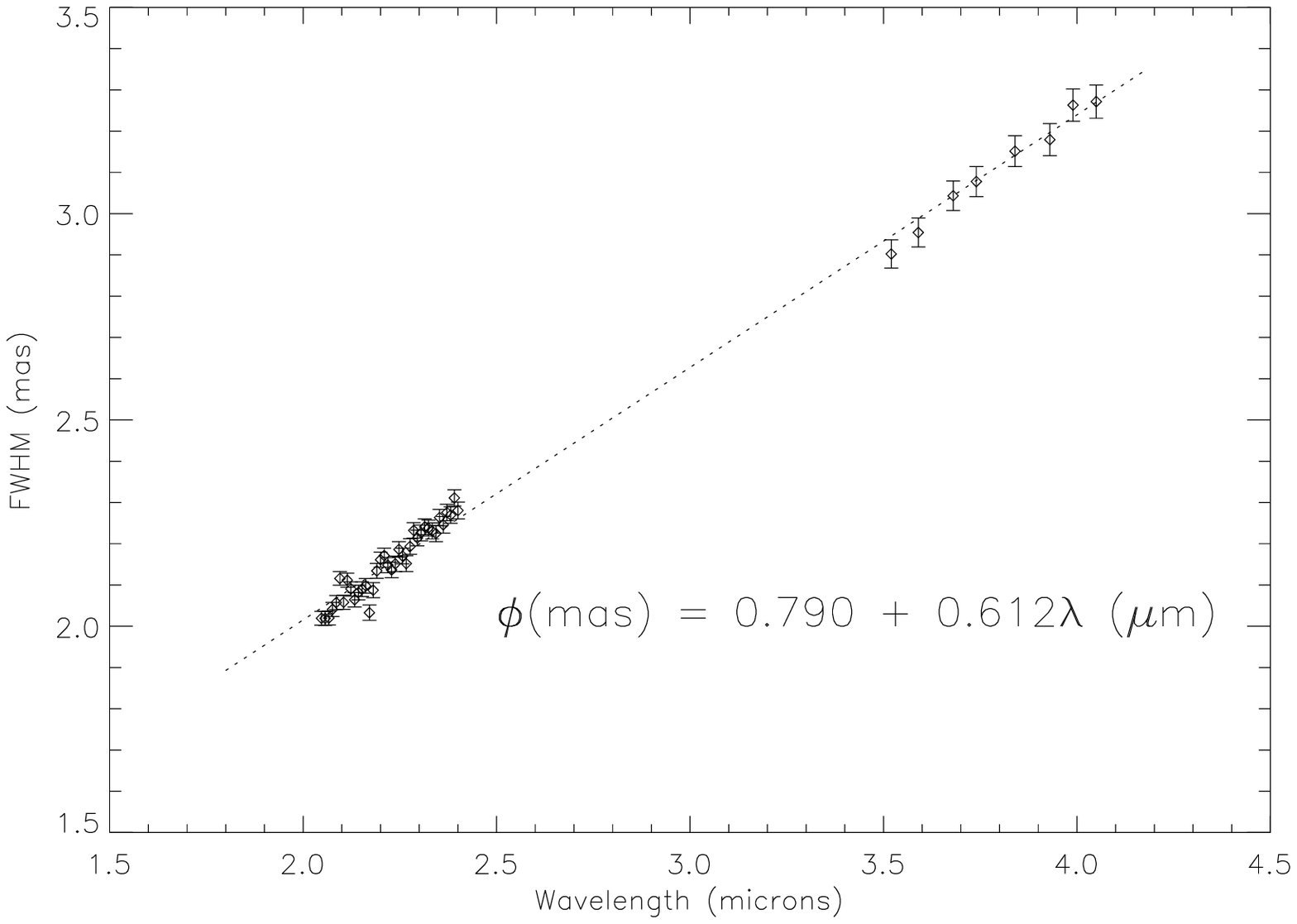}
\caption{Wavelength-dependent linear fits. {\bf Left:} The derived uniform-disk angular sizes as a function of wavelength are shown in diamond symbol along with error bars. A linear fit to these sizes (dotted line) is also shown here. The dip seen at around 2.17$\mu$m is due to the presence of compact Br $\gamma$ emission line (see Section 4 for details). {\bf Middle:} Same as the left figure, but a zoomed view of the Br $\gamma$ emission line region. {\bf Right:} Same as the left figure except for a Gaussian distribution.
}
\label{ud}
\end{figure}

As the wavelength-independent size models fail to fit the measurements, we fit the data with wavelength-dependent sizes. The results of these geometrical model fits to the measured data points are shown in Figure~\ref{ud}.

The observed wavelength dependence in the 2-4 $\mu$m region has a simple linear relationship to first order. The parameters of a linear fit to the derived apparent wavelength dependent diameters are given in Figure~\ref{ud}. The $\chi^2_R$ value for this model fit is 2.1. The measured uniform-disk diameter in the center of the L-band ($\theta_L$ = 5.04 mas) is $\sim$ 44\% larger than in the center of the K-band ($\theta_K$ = 3.49 mas). The steep slopes of these linear relations suggest that the 2-4$\mu$m emission source must be extended with strong radial temperature dependence. Earlier multi-color (near- and mid-infrared) interferometry of two YSO disks have shown similar wavelength dependence \citep{2008A&A...485..209A, 2008ApJ...676..490K}. 
In the following section we explore more complex extended disk models to explain our observations.


\subsection{Geometrically thin, optically thick accretion disk models}
We compared our interferometric data with an accretion disk model by Hillenbrand et al. (1992)
based on an SED analysis. 
In this classical accretion disk model, the temperature distribution of dust is derived by combining the contributions from a radiatively heated reprocessing component T$_{rep}$ and a viscously heated accretion component T$_{acc}$. The parameters of the disk are the following: The inner disk (hole) radius is R$_{in}$ = 0.22 AU (10 R$_*$), the outer disk radius is 62 AU and the mass accretion rate ($\dot{M}$) is 1.98 $\times$ 10$^{-5}$ M$_\odot$ yr$^{-1}$. The stellar radius is 4.8R$_\odot$, the stellar effective temperature is 11,220K, the distance to the star is 650pc and the stellar mass is 5.3M$_\odot$. The inner disk temperature is 2480K and the outer disk temperature is 40K. The resulting flat accretion disk model highly overestimates $V^2$ in the K- and L-bands. We get a reduced-chi-square ($\chi^2_R$) value of 664 for the interferometric data. The model results are given in Table 1. The physical reason that the classical accretion disk model fails is that it creates too much flux in the inner region of the disk because of the added accretion luminosity, making the angular size much smaller than observed.

\begin{figure}[bthp]
\centering
\includegraphics[width=0.98\hsize]{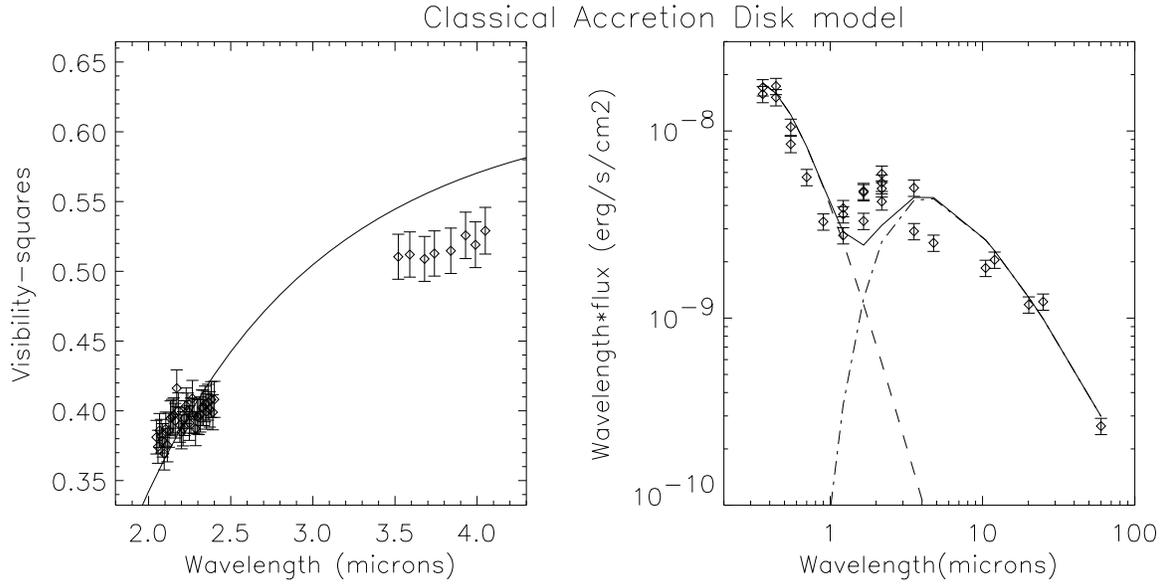}
\caption{Classical accretion disk model fits. Interferometric data points are shown with error bars. Shown in continuous line is an accretion disk model with the inner radius of the disk (hole size) treated as a free model parameter. The outer radius of the disk is fixed at 50 AU. {\bf Right:} Photometric data taken from the literature are shown along with the SED model for the same accretion disk model. The dashed line shows the blackbody spectral distribution of the central star and the dashed-dotted line shows the blackbody emission from the disk.
}
\label{dust}
\end{figure}

We have also fit visibility data with a similar accretion disk model (Figure~\ref{dust}) by treating the inner disk (hole) radius as a free model parameter. In addition, the mass accretion rate was increased by a factor of 1.5 (i.e. 2.97 $\times$ 10$^{-5}$ M$_{\odot}$ yr$^{-1}$) in order to fit SED data. 
The $\chi^2_R$ value for this model fit to interferometric measurements is 2.59. The derived inner disk (hole) radius is 1.66 $\pm$ 0.01 mas (0.54AU; $\sim$24R$_*$) and the inner disk temperature is 1457K. The outer disk radius is fixed at 50 AU  where the disk temperature is 51K. This model fits the K band data well, but is about 3 $\sigma$ too high in $V^2$ compared to the L band data. The extremely large accretion luminosity required by the SED data poses challenges to these two accretion disk models. Thus the classical model that could fit the SED fails here, just like in the previous interferometric studies of Herbig Ae/Be disks \citep{Millan-Gabet01,2004ApJ...613.1049E}. 
Moreover, it is not sufficient to adjust the inner dust radius and temperature alone. A modification of the disk temperature profile is needed as in the case of more recent studies \citep{2007ApJ...657..347E, 2008A&A...485..209A, 2008ApJ...676..490K}; this is discussed in the following section.

\subsection{Geometrically thin, optically thick disk with power-law temperature distribution}
\begin{figure}[bthp]
\centering
\includegraphics[width=0.98\hsize]{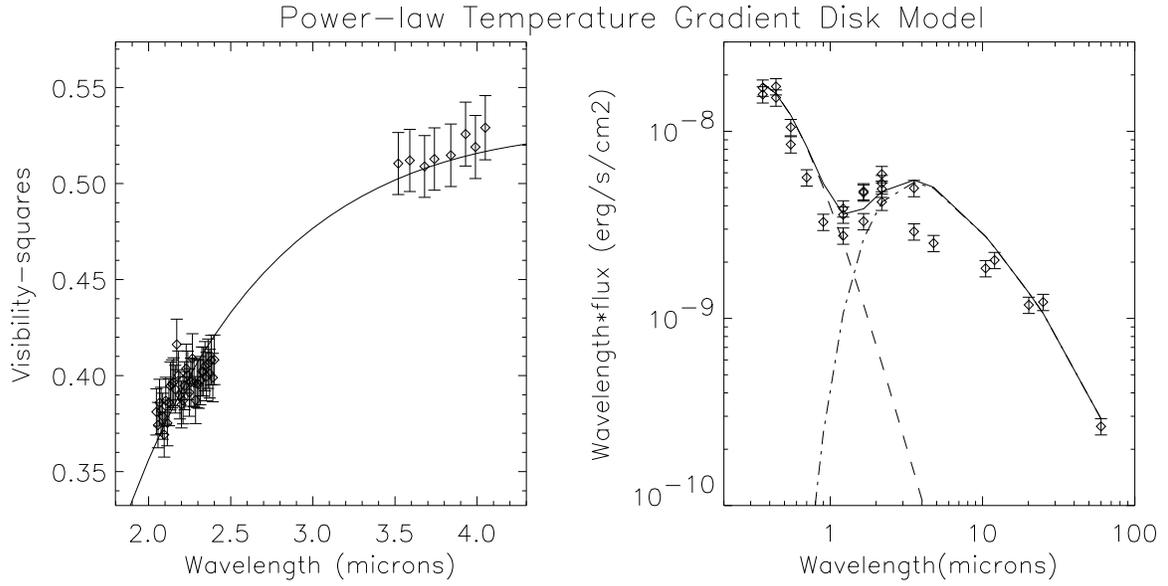}
\caption{Power-law disk model fits. {\bf Left:} Interferometric data points are shown with error bars. The solid line is a power-law temperature gradient disk model. The power-law parameter, the inner disk radius and the inner disk temperature are free parameters. The outer radius is fixed at 20AU. {\bf Right:} Photometric data taken from the literature are shown with error bars. The solid line is the SED model for the same accretion disk model. The dashed and dashed-dotted lines are the SEDs of the star and disk.
}
\label{power_law}
\end{figure}

\begin{figure}[bthp]
\centering
\includegraphics[width=0.475\hsize]{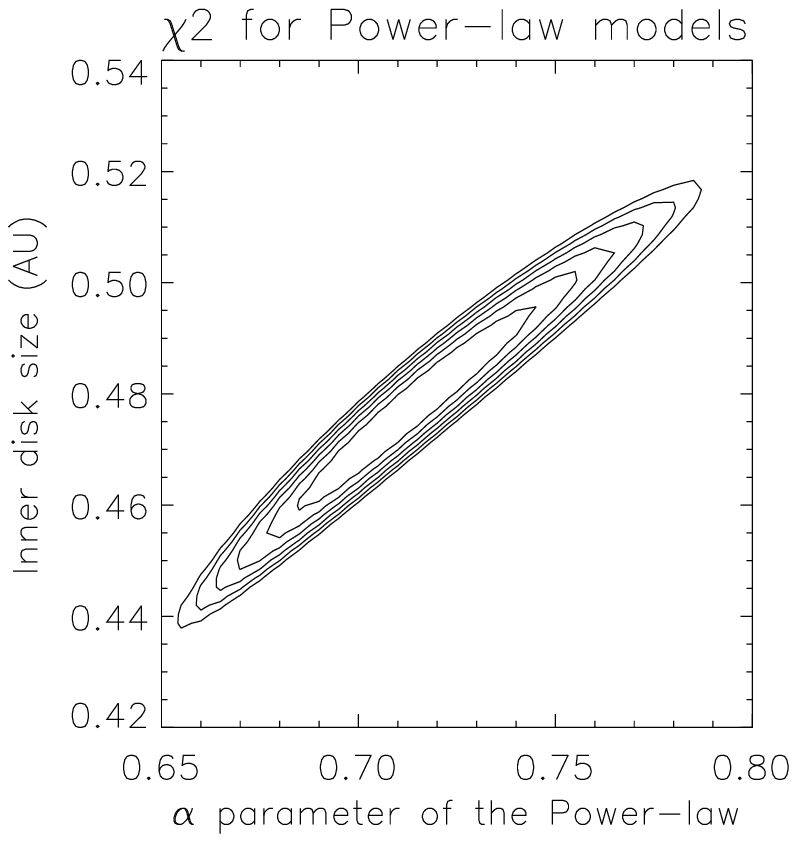}
\caption{Contour map of $\chi^2_R$ as a function of the power-law parameter $\alpha$ and the inner disk radius. The contour lines - from inner to the outer - refer to $\chi^2_R$ values of 1.0, 1.2, 1.4, 1.6, 1.8 and 2.0.
}
\label{chi_sq}
\end{figure}

The wavelength dependence of the measured visibilities and SED data shown above implies an inverse $T(r)$ relationship. We fit our data with a simple model with a power-law temperature gradient of functional form $T(r)$ $\propto$ $r^{-\alpha}$, where $r$ is the radial distance from the central star and $\alpha$ is the power-law parameter. The model $V^2$ is derived using Eqn.~\ref{vdisk} and Eqn.~\ref{fdisk} with a power-law temperature distribution. The radius of the inner disk and $\alpha$ are treated as free parameters. The temperature of the inner disk is fixed at 1800K and the optical depth at 1 for the entire disk, in order to satisfy the SED data; a larger value for the optical depth overestimates the disk flux. The resultant model fits are shown in Figure~\ref{power_law}. The derived inner disk radius is 1.47 $\pm$ 0.02 mas (0.477AU; $\sim$ 21R$_*$) and the derived value for $\alpha$ is 0.71 $\pm$ 0.01. The outer radius is fixed at 20AU (in order for the SED to be consistent with the far-infrared photometric data) and the corresponding temperature is 125K.
The $\chi^2_R$ for this model fit to our interferometric data is 0.73, and the sensitivity of $\chi^2_R$ to the model parameters is shown in Figure~\ref{chi_sq}. 
As the power-law disk model is a satisfactory fit, we consider it the simplest model that explains the data.  However, one can also conceive of more complex and physically plausible models such as optically thin inner disk regions (holes) surrounded by optically thick outer disks, perhaps including puffed up inner rims.  Our initial exploration of such models suggests that they can also fit our data, but the interferometric data themselves do not drive us to such complex models.

\begin{table}[h]
\begin{center}
\begin{tabular}{lccccr}
\hline
\hline
Model                    & R$_{in}$       & T$_{in}$ &$\alpha$& $\dot{M}$&$\chi_{R}^2$\\
                         & (AU)           & (K)        &        & ($\times$ 10$^{-5}$ M$_\odot$ yr$^{-1}$)&\\ 
\hline
Classical accretion disk & 0.22           & 2480    & 0.75         &1.98 & 664   \\
(parameters fixed at published values)       & & & & & \\
Classical accretion disk & 0.538$\pm$0.001& 1457    & 0.75         &2.97 &2.59   \\
(R$_{in}$ and $\dot{M}$ varied)       & & & & & \\
Modified power-law disk           & 0.477$\pm$0.005& 1800    & 0.71$\pm$0.01&-    &0.73  \\
\hline
\end{tabular}
\label{resultsTable}
\caption{Derived parameters for the disks models presented in Sections 3.2 and 3.3. The $\chi_{R}^2$ values are for the interferometer data (only).}
\end{center}
\end{table}

\subsection{Binary star model}
In this section we investigate whether the data can be explained, without invoking hot/warm circumstellar dust, by the effect of a companion star.
The visibility model for a binary system can be written as 

\begin{equation}
V_{binary}(\lambda) = \frac{\sqrt{V_p^2+R^2 ~V_s^2+2 ~|V_p| ~|V_s| ~R ~cos((2\pi/\lambda)~{\bf B \cdot s})}}{1+R}
\label{binary_eqn}
\end{equation}
where R is secondary-to-primary flux ratio, $V_p$ and $V_s$ are visibilities of primary and secondary components, and {\bf s} is the binary separation.

\begin{figure}[bthp]
\centering
\includegraphics[width=0.475\hsize]{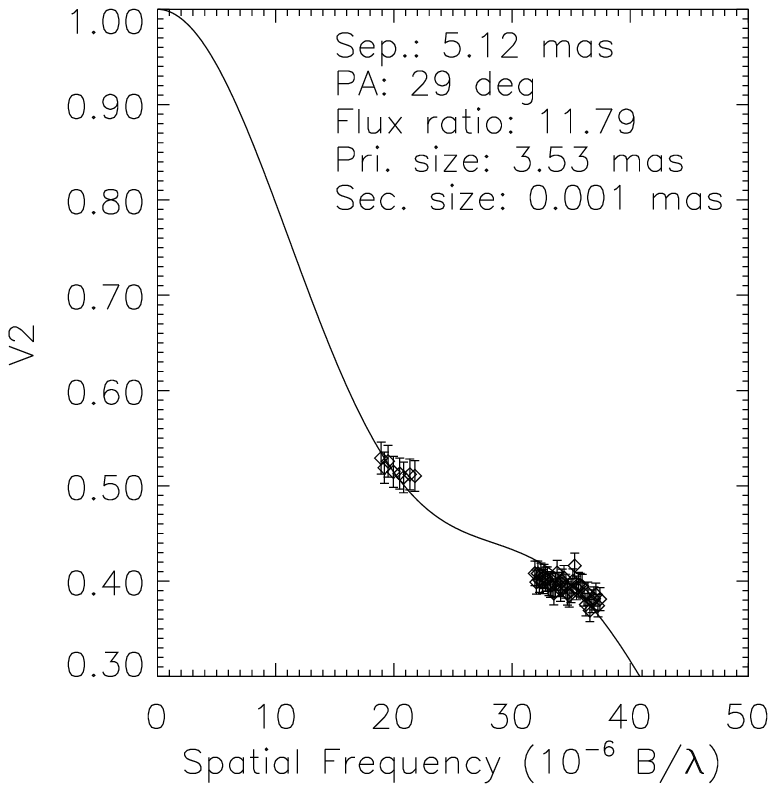}
\includegraphics[width=0.475\hsize]{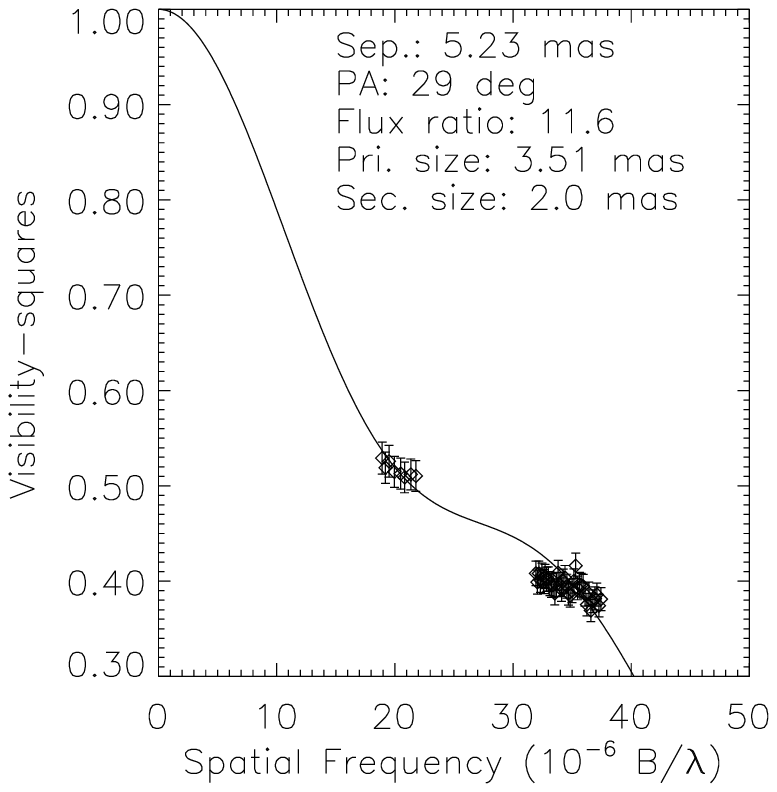}
\caption{Binary model fits. {\bf Left:} The multi-wavelength measurements are plotted against spatial frequency. Also shown is the best fit binary model (continuous line) assuming an unresolved companion. {\bf Right:} Same as the previous caption but for a companion of 2 mas diameter.
}
\label{binary}
\end{figure}

The derived parameters from a binary model fit to our measurements are the following: projected binary separation along the position angle of 29$^{\rm o}$ (east of north) = 5.12$\pm$0.01mas, primary-to-secondary flux ratio = 11.79$\pm$0.30, and uniform-disk angular diameter of the primary component (YSO disk size) = 3.53$\pm$0.01 mas (a binary model with two
point sources fails to fit SED data). The diameter of the secondary component is fixed at 0.001 mas (effectively unresolved). The model fit to the data is shown in Figure~\ref{binary}. The $\chi^2_R$ value is 0.94. In this scenario, the secondary component would have the same K-L color of 1.34 as the primary YSO disk suggesting that the potential companion were a YSO disk with significant IR excess.
In this case, our observations would have resolved the disk around the companion. Hence, we also fit our measurements by fixing the companion size to be 2 mas in order to explore this possibility. The derived parameters for this binary model are the following: projected binary separation along the position angle of 29$^{\rm o}$ (east of north) = 5.23$\pm$0.01mas, primary-to-secondary flux ratio = 11.6$\pm$0.3, and uniform-disk angular diameter of the primary component (YSO disk size) = 3.51$\pm$0.01 mas. The model fit to the data is also shown in Figure~\ref{binary}. The $\chi^2_R$ value is 1.57, but additional observations at other baseline orientations would provide significant additional constraints on the binary model. The derived flux ratio of $\sim$ 12 suggests a cooler stellar companion with a smaller inner disk radius compared to the primary star. However, such a companion model would overestimate SED at the shorter wavelengths because of the flux from the companion photosphere. Hence, a binary scenario is not favored as an explanation for our measurements. 

\section{Discussion}

\begin{figure}[bthp]
\centering
\includegraphics[width=0.5\hsize]{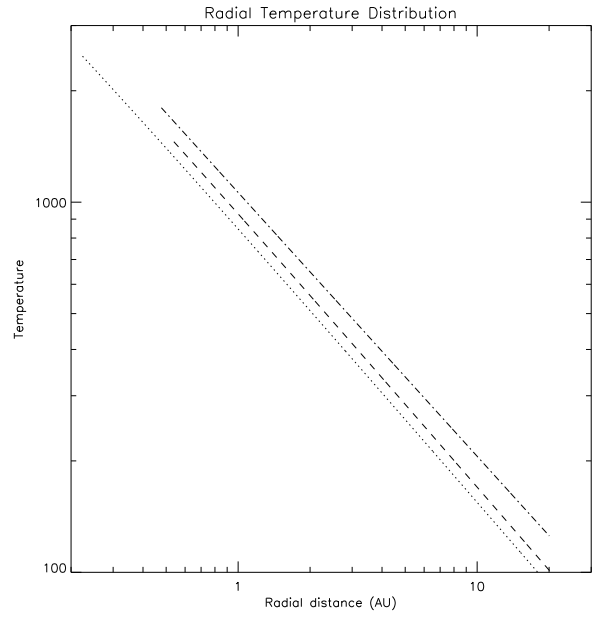}
\caption{The radial temperature profiles of various disk models. The dotted and dashed lines refer to the two cases of the classical-accretion disk model (Section 3.2; Table 1). The dotted-dashed line refers to the power-law temperature gradient disk model (Section 3.3; Table 1).
}
\label{temp_profile}
\end{figure}

The KI measurements presented here agree with previous observations from PTI of the
K-band size of MWC 419 \cite{2003Ap&SS.286..145W}. The addition of the spectrally
dispersed K-band information and new L-band observations, also spectrally dispersed, provide powerful new constraints
on the physical structure of the material surrounding the central star.
As shown the left-most panel of Figure 2, the wavelength dependence of
the size is apparent within both the K and L bands; the combination
of the two bands allows an even tighter constraint on the temperature
power-law than possible using either data set alone.  
The radial temperature profiles of the disk models presented in
Sections 3.2 and 3.3 are shown in Figure~\ref{temp_profile}. The
temperature profile of the best fit power-law model is very
similar to that of a classical, geometrically thin accretion/reprocessing disk ($\alpha = -0.75$).

As discussed earlier (Section 3.2), the accretion disk model requires
an unrealistically large accretion rate, on the order of a few times
10$^{-5}$ M$_\odot$ yr$^{-1}$ in order to fit the SED data. Simple estimates of the accretion rate
for this star range from 5$\times$10$^{-7}$ to 10$^{-6}$ M$_\odot$
yr$^{-1}$ based on mass loss data and upper limits to radio emission
at 3.6 cm \citep{2007Ap.....50..207K, 1995A&A...301..155B,
1993ApJS...87..217S}. The necessity of an unrealistically large
accretion rate in the accretion disk model is probably an artifact of the physically unrealistic
assumption of a zero-thickness disk which causes the re-processing of radiation to be very inefficient. Hence, we do not favor an
accretion disk model for the MWC~419 disk.

A power law temperature profile with radius in which the radial 
exponent is a free parameter is a very good fit to the data,
because such a simple model roughly accounts for the
vertical scale height as a function of radius. 
The derived slope of -0.71 is 
consistent with a ``flat" disk geometry that one tends to find for Herbig Be stars
\citep{2008A&A...485..209A, 2008ApJ...676..490K}.

Earlier studies \citep{2002ApJ...579..694M, 2004ApJ...613.1049E}
using broadband, usually single-wavelength, interferometric data, 
recognized a difference in the near-infrared size vs.
luminosity behavior of high luminosity objects (pre-main sequence Be) compared to 
lower luminosity ones (pre-main sequence Ae), the former being more consistent with 
``classical disk" models.  This has been revisited most recently by 
Vinkovi{\'c} \& Jurki{\'c} (2007), who use a model-independent comparison
of visibility to scaled baseline and find a distinction between
low-luminosity ($\lesssim$ 10$^3$ L$_ \odot$) and high-luminosity
($\gtrsim$ 10$^3$ L$_\odot$) YSO disks where the luminosity break point
corresponds to approximate spectral type B3-B5.  These authors then
modelled this observable with a ring and halo model for low luminosity Herbigs,
a halo alone for T Tauris and an accretion disk for high luminosity Herbigs.

However, multi-wavelength interferometric studies have not always
supported these conclusions when objects are modeled in detail.
Kraus et al. (2008) observed the B6 star MWC 147 in H, K and N bands.  They
performed Monte Carlo modeling and found that the interferometric and 
SED date were not well fit with a standard, irradiated accretion
disk alone, but were well fit with the standard disk plus emission
from hot, optically thick gas within the innermost radius of the dust disk. In
contrast, 
Acke et al. (2008) observed the B1.5 star MWC 297 in H, K and N,
and found that they could not fit the data with a single accretion
disk, even with the radial temperature exponent as a free parameter.
Instead, they used a three component geometric model, with
characteristic blackbody temperatures of 1700, 920 and 520 K.  Our
results show that MWC~419 (B8, 330 L$_\odot$) has the disk
characteristics of a high-luminosity object in the categories of
Monnier \& Millan-Gabet (2002)
but a luminosity lower than the
break point of 10$^3$  L$_\odot$ identified by 
Vinkovi{\'c} \& Jurki{\'c} (2007). Examining Figure 2 of Vinkovi{\'c} \& Jurki{\'c} (2007) clearly shows 
MWC~419 fits well within the population
of lower luminosity Herbigs in their model-independent comparison.
Their physical interpretation of the low-luminosity Herbig group
is an optically thick disk with an optically thin dust sublimation
cavity and an optically thin dusty outflow.  Our multi-wavelength
results do not support grouping MWC 419 in that physical model, 
but perhaps incorporating multi-wavelength data in the model-independent
visibility groupings would have produced a different result.
The new multi-wavelength data set presented here provides significant
new information to aid in determining the physical conditions
of these young stars, and our data show that the disk surrounding the B8 star
MWC 419 is closer in physical characteristics to the
more massive Be stars than to the Herbig Ae and T Tauri
stars.

Another result from the spectrally dispersed data is that the measured
$V^2$ is marginally higher at 2.17$\mu$m than at 
adjacent wavelengths.  This 
$V^2$ feature is seen in all 4 independent data sets.
As this is the wavelength of Br $\gamma$, and  
Brackett recombination emission lines such as
Br $\gamma$ and Br 10-20 are observed in the spectrum of
MWC~419 \citep{1984PASP...96..297H}, 
we attribute the difference in 
line and continuum visibilities to a more compact Br $\gamma$ emitting region. 
Similar behavior in the apparent size in the Br $\gamma$ line
compared to the neighboring continuum has been seen before in
spectrally resolved interferometry from both VLTI and KI
\citep{2008ApJ...676..490K, 2009ApJ...692..309E}. The derived
uniform-disk diameter for the Br $\gamma$ emitting region is 3.33$\pm$0.03
mas, which is $\sim$ 4\% less than the continuum size (3.46 mas) around
Br $\gamma$ (Figure~\ref{ud}). The significance level of this
detection of the Br $\gamma$ emitting region with respect to the
continuum region is $\sim$ 2 $\sigma$.  The actual size of the Br
$\gamma$ emission line region could be much smaller than the value
reported here because of the coarse spectral resolution of our
measurements. Detection of Br $\gamma$ emission inside the innermost
dust radius suggests that the
disk is optically thin in the inner region where atomic hydrogen gas exists.

\section{Summary}
This article reports the first milliarcsecond angular resolution observations of a HAeBe star (MWC~419) 
providing L-band, as well as simultaneously obtained K-band data, both spectrally dispersed. This multi-wavelength observational capability is well suited to probing the temperature distribution in the inner regions of YSO disks, which is very important for distinguishing  
between models and gaining insight into the three dimensional geometry of the inner disk.
Such measurements could distinguish discrete spatial distributions, such as dust-rims, from relatively-smooth spatial distributions, such as classical accretion disks, based on their distinct wavelength dependent disk sizes. In addition, interferometric measurements in the relatively unexplored L-band provide needed constraints to the disk/envelope geometry and temperature structure.

Simple geometrical pole-on disk models are used to infer a linear relationship between the derived object size and wavelength in the 2-4 $\mu$m region, suggesting a simple physical model for the disk. The steep slopes of these linear relations imply that the disk is extended with a radial temperature gradient.
We find that the accretion disk model of Hillenbrand (1992)
derived from SED analysis does not fit our interferometric measurements. An updated accretion disk model with 
accretion rate 1.5 times larger and inner cavity 2.4 times larger
fits the K band data well, but lies 3$\sigma$ above the L band data. 
However, both of these classical accretion disk models predict an unrealistically large accretion rate of $\sim$ 3$\times$10$^{-5}$ M$_{\odot}$ yr$^{-1}$ to fit the SED data. A power law temperature profile with a slightly shallower slope of -0.71 fits both the spectrally dispersed interferometric measurements and the SED satisfactorily, suggesting a relatively flat disk geometry for MWC~419. 
The measured disk size at Br $\gamma$ reveals the presence of compact 
emitting hydrogen gas in the inner regions of the disk. 
A more complete sample of YSO disk observations with adequate wavelength and (u,v) coverage, plus detailed radiative transfer modeling, are required to address the intriguing inner disk geometry in these sources 

\section*{Acknowledgment}
Keck Interferometer is funded by the National Aeronautics and Space Administration (NASA). Observations presented were obtained at the W. M. Keck Observatory, which is operated as a scientific partnership among the California Institute of Technology, the University of California, and NASA.  The Observatory was made possible by the generous financial support of the W. M. Keck Foundation.  
We thank E. Appleby, B. Berkey, A. Booth, A. Cooper, S. Crawford, W. Dahl, C. Felizardo, J. Garcia-Gathright, J. Herstein, R. Ligon, D. Medeiros, D. Morrison, T. Panteleeva, B. Smith, K. Summers, K. Tsubota, G. Vasisht, E. Wetherell, for their contributions to the instrument development, integration and operations. S. Ragland also thanks M. Hrynevych, M. Kassis and J. Woillez for useful discussions. We thank the referee for constructively critical comments that have helped us to significantly improve the paper. 



\end{document}